\documentclass[conference]{IEEEtran}
\IEEEoverridecommandlockouts

\usepackage{cite}
\usepackage{amsmath,amssymb,amsfonts}
\usepackage{algorithmic}
\usepackage{graphicx}
\usepackage{textcomp}
\usepackage{xcolor}
\def\BibTeX{{\rm B\kern-.05em{\sc i\kern-.025em b}\kern-.08em
    T\kern-.1667em\lower.7ex\hbox{E}\kern-.125emX}}
\usepackage{subcaption}
\usepackage{dblfloatfix}
\newcommand{\linebreakand}{%
  \end{@IEEEauthorhalign}
  \hfill\mbox{}\par
  \mbox{}\hfill\begin{@IEEEauthorhalign}
}

\begin{document}

\title{Active Inference for Closed-loop transmit beamsteering in Fetal Doppler Ultrasound\\
\thanks{This project is supported by the Chips Joint Undertaking (Grant Agreement No. 101095792) and its members Finland, Germany, Ireland, the Netherlands, Sweden, Switzerland. This work includes top-up funding from the Swiss State Secretariat for Education, Research and Innovation.
}
}

\author{\IEEEauthorblockN{Beatrice Federici}
\IEEEauthorblockA{Department of Electrical Engineering\\
	Eindhoven University of Technology\\
	Eindhoven, The Netherlands\\
	\texttt{b.federici@tue.nl}}
\and
\IEEEauthorblockN{Ruud JG van Sloun}
\IEEEauthorblockA{Department of Electrical Engineering\\
	Eindhoven University of Technology\\
	Eindhoven, The Netherlands\\
	\texttt{r.j.g.v.sloun@tue.nl}}
\and
\IEEEauthorblockN{Massimo Mischi}
\IEEEauthorblockA{Department of Electrical Engineering\\
	Eindhoven University of Technology\\
	Eindhoven, The Netherlands\\
	\texttt{m.mischi@tue.nl}}
}

\maketitle

\begin{abstract}
Doppler ultrasound is widely used to monitor fetal heart rate during labor and pregnancy. Unfortunately, it is highly sensitive to fetal and maternal movements, which can cause the displacement of the fetal heart with respect to the ultrasound beam, in turn reducing the Doppler signal-to-noise ratio and leading to erratic, noisy, or missing heart rate readings. To tackle this issue, we augment the conventional Doppler ultrasound system with a rational agent that autonomously steers the ultrasound beam to track the position of the fetal heart. The proposed cognitive ultrasound system leverages a sequential Monte Carlo method to infer the fetal heart position from the power Doppler signal, and employs a greedy information-seeking criterion to select the steering angle that minimizes the positional uncertainty for future timesteps. The fetal heart rate is then calculated using the Doppler signal at the estimated fetal heart position.
Our results show that the system can accurately track the fetal heart position across challenging signal-to-noise ratio scenarios, mainly thanks to its dynamic transmit beam steering capability. Additionally, we find that optimizing the transmit beamsteering to minimize positional uncertainty also optimizes downstream heart rate estimation performance.
In conclusion, this work showcases the power of closed-loop cognitive ultrasound in boosting the capabilities of traditional systems.
\end{abstract}

\begin{IEEEkeywords}
action-perception, cognitive systems, Doppler ultrasound
\end{IEEEkeywords}
\section{Introduction}
\noindent Fetal heart rate (HR) monitoring is critical to adequate prenatal care, providing essential insight into the health and well-being of the fetus. One of the most widely used and non-invasive methods for monitoring fetal HR is Doppler ultrasound \cite{hamelmann2019doppler}. In fetal Doppler ultrasound, a transducer is carefully placed on the maternal abdomen to insonify the fetal heart using an ultrasound beam. The backscattered echoes are then recorded and processed to detect motion (i.e. Doppler processing), from which the HR is derived. 

Despite its extensive use, Doppler ultrasound is highly sensitive to both fetal and maternal movements, which can cause the displacement of the fetal heart relative to the ultrasonic beam \cite{avalon}. When this happens, the signal-to-noise ratio (SNR) of the Doppler signal drops significantly, resulting into 
erratic and noisy HR traces or no HR tracing at all. Sometimes, maternal insertion can also occur, where the system may pick up and display the maternal HR \cite{avalon}. 
During prolonged periods of such weak and mixed signals, the fetus is not being adequately monitored, and the clinical staff has to reposition the ultrasound transducer.

To improve robustness to displacement, modern ultrasound transducers for fetal HR monitoring often incorporate large elements that generate relatively wide beams (e.g., an effective radiating area at -12~dB of 4.47 $\pm$ 0.89 cm$^{2}$ \cite{avalon}). While this approach helps mitigate the impact of small displacements, larger displacements such as fetal reorientation still compromise the acquisition. Additionally, the wider beam has a lower beamforming gain compared to tightly focused beams, resulting in reduced resiliance to noise. Transducer design therefore constitutes a trade-off between the achieved Dopper SNR and its robustness to fetal movements.  

In this work, we present an alternative approach. Instead of widening the ultrasound beam to reduce the sensitivity to displacement, we propose a closed-loop cognitive ultrasound system \cite{10689436} that autonomously steers a highly focused ultrasound beam to ensure that the heart is always properly insonified, tracking displacements relative to the ultrasound transducer. 
In practice, we adopt a phased array and control it through a rational agent whose task is to sequentially select the optimal focused transmit beam for tracking the position of the fetal heart. To this end, the agent combines a sequential Monte Carlo method (particle filter) to infer a Bayesian posterior fetal heart position from the measured power Doppler signal at timepoint $t$, with an information maximization strategy that selects the future beam steering angle that is expected to reduce the positional uncertainty the most at timepoint $t+1$. The one-dimensional Doppler signal at the estimated heart position is then used to compute the HR.

The remainder of this paper is organized as follows. Section~\ref{sec:methods} describes the proposed Doppler ultrasound system, including the closed-loop probabilistic framework and the HR estimation approach. 
Section~\ref{sec:experiments} presents the experiments conducted to evaluate the performance of the system \textit{in-silico} and \textit{in-vitro}. We discuss the results in Section~\ref{sec:discussion} and conclude in Section~\ref{sec:conclusion}.




\begin{figure}[t!]
\centering
\includegraphics[width=0.5\textwidth]{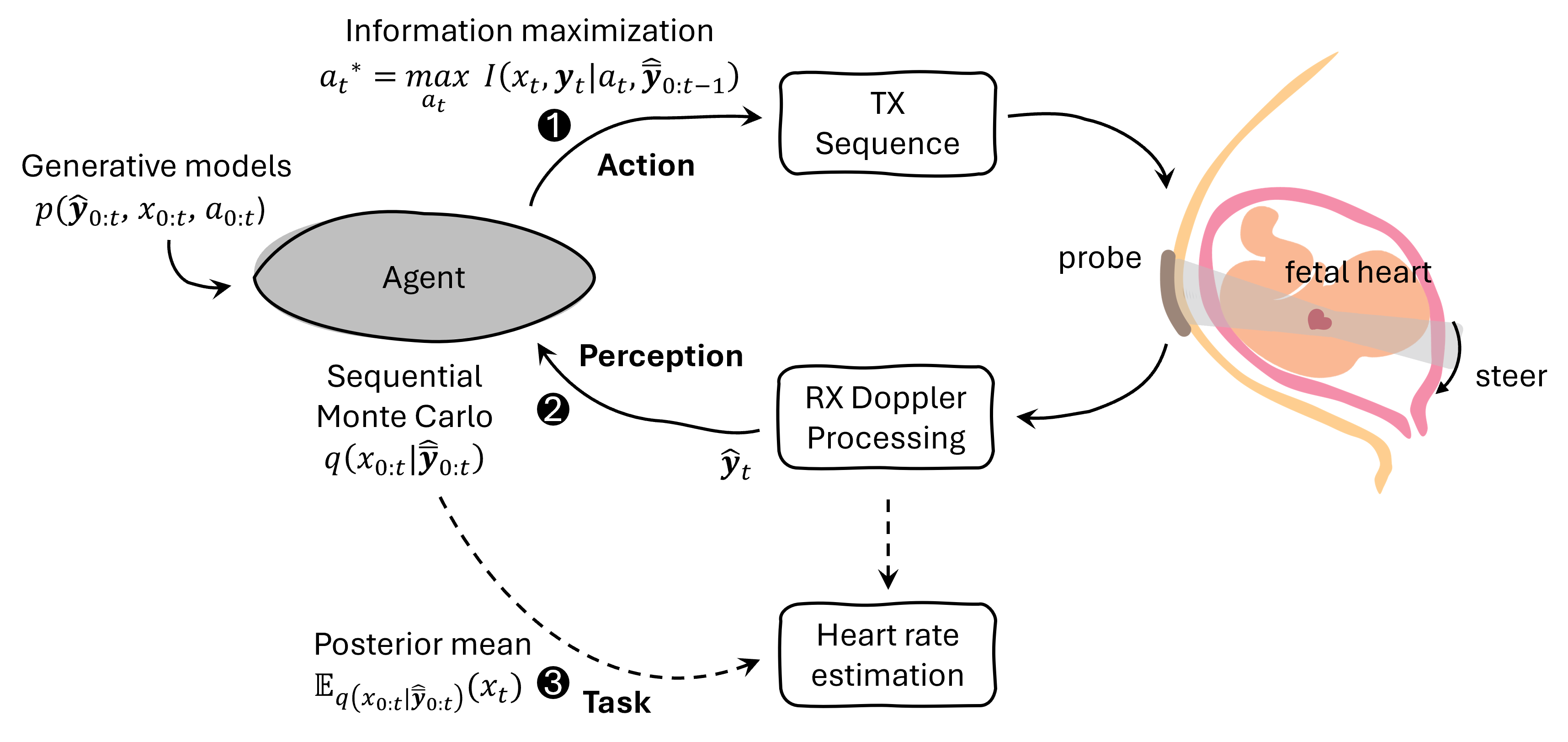}
\caption[Proposed closed-loop fetal Doppler ultrasound.]{Closed-loop fetal Doppler ultrasound.}
\label{fig:scheme}
\end{figure}

\section{Methods}
\label{sec:methods}
\noindent We model the fetal heart as a moving Doppler target, with unknown fetal HR and position. This target is observed by a phased array transducer, which probes the external environment using a pulsed wave Doppler scheme with a focused beam having a fixed focal depth and an adaptive steering angle. At any time point $t$, a rational agent is tasked with the selection of the optimal future steering angle, $a^{*}_{t+1} \in \boldsymbol{\Theta}^{tx}$, that maximizes information gain with respect to the future state $x_{t+1}$: the angular position of the Doppler target at timestep $t+1$. This steering action subsequently leads to a new observation $\hat{\textbf{y}}_{t+1} \in \mathbb{R}^{N_{\Theta}}$, a clutter-filtered power Doppler image beamformed on a polar grid $(\boldsymbol{\Theta}, \textbf{r})$ and subsequently integrated across the depth. 

Figure~\ref{fig:scheme} illustrates the proposed closed-loop cognitive ultrasound system. It comprises a probabilistic generative model and iterates between two steps: (1) perception, the updating of the Bayesian posterior about fetal heart position, and (2) action, the selection of an information-maximizing transmit steering angle.



\subsection{Generative Model}
The agent's generative model across Doppler observations ($\textbf{y}_{0:T}$), angular fetal heart positions ($x_{0:T}$), and beamsteering actions ($a_{0:T}$) is given by:
\begin{equation}
    p(\textbf{y}_{0:T}, x_{0:T}, a_{0:T}) = p(\textbf{y}_{0:T}|x_{0:T}, a_{0:T})p(a_{0:T}|x_{0:T})p(x_{0:T}).
\end{equation}
We assume that each measurement $\textbf{y}_{t}$ is conditionally independent of measurements at other time points given $x_{t}$ and $a_{t}$ and that the state dynamics can be described as a first-order Markovian process. Hence, the state dynamics and the measurement process are governed by the state transition $p(x_{t} | x_{t-1})$ and the observation model $p(\textbf{y}_{t} | x_{t}, a_{t})$, respectively. We adopt a linear Gaussian state transition dynamics $p(x_{t} | x_{t-1}) =  \mathcal{N}(x_{t-1}, \sigma_{x})$, and a Gaussian non-linear observation model $p(\textbf{y}|x_{t}, a_{t}) = \mathcal{N}(f_{y}(x_{t};a_{t}), \Sigma_{y})$, where $f_{y}(\cdot;a_{t}): \mathbb{R}^{1}\rightarrow \mathbb{R}^{N_{\Theta}}$ maps the angular position of the target to a simulated power Doppler measurement, considering an approximate action-conditional transmit beam profile:
\begin{equation}
f_{y}(x_t;a_{t}) = e^{-\frac{(\boldsymbol{\Theta}-a_t)^2}{2bw}} e^{-\frac{(\boldsymbol{\Theta}-x_t)^2}{2hw}}.
\end{equation}

\subsection{Perception}
Since our observation model $p(\textbf{y}_{t} | x_{t}, a_{t})$ is highly nonlinear and high-dimensional, we use approximate Bayesian inference to update the agent's beliefs about the state given new observations and actions $\hat{\bar{\textbf{y}}}_{0:t} =  \{\hat{\textbf{y}}_{0:t}, a_{0:t}\}$. More specifically, we use a particle-based approximation to the posterior $q(x_{0:t}|\hat{\bar{\textbf{y}}}_{0:t})$
with a set of $N_p$ samples/particles and importance weights $\{x^{i}_{t}, w^{i}_{t}\}_{i=1}^{N_{p}}$: 
\begin{equation}
q(x_{0:t}|\hat{\bar{\textbf{y}}}_{0:t}) = \sum_{i=1}^{N_{p}}  w_{t}^{i} \delta (x_{t} - x^{i}_{t})~,
\end{equation}
where we track the particles and weights using a sequential Monte Carlo filtering method, the \textit{particle filter}.

\subsection{Action}

Given the updated posterior, the agent selects the beam steering angle, $a^{*}_{t+1}$, that minimizes the expected posterior entropy. This is equivalent to selecting the action that maximizes the conditional mutual information between the state and future observations \cite{lindley1956measure}:
\begin{equation}
\begin{split}
a_{t+1}^{*} &= \max_{a_{t+1}\in \boldsymbol{\Theta}^{tx}} I(x_{t+1}, \textbf{y}_{t+1} | a_{t+1}, \hat{\bar{\textbf{y}}}_{0:t}) \\&=  \max_{a_{t+1}\in \boldsymbol{\Theta}^{tx}} H(\textbf{y}_{t+1} | a_{t+1}, \hat{\bar{\textbf{y}}}_{0:t}) - H(\textbf{y}_{t+1} |x_{t+1}, a_{t+1}, \hat{\bar{\textbf{y}}}_{0:t}) ~. 
\end{split}
\label{eq:3}
\end{equation}
This reduces to maximizing the marginal differential entropy of future observations as the uncertainty about $\textbf{y}_{t+1}$ given $x_{t}$ is independent of the action $a_{t+1}$ according to the defined generative model $p(\hat{\textbf{y}}_{t} | x_{t}, a_{t}) \sim \mathcal{N}(f_{y}(x_{t}, a_{t}); \sigma_{y})$.

In practice, the agent uses the set of particles $\{x^{i}_{t}, w^{i}_{t}\}_{i=1}^{N_{p}}$ to derive hypothetical futures states $x_{t+1}^{i} \sim p(x_{t+1}^{i}|x_{t}^{i})$, using the linear Gaussian state transition model. Then, for each possible action $a_{t+1}$, the agent computes $N_{p}$ hypothetical measurements $\{\textbf{y}_{t+1}^{i}|a_{t+1} \sim p(\textbf{y}_{t+1}|x^{i}_{t+1}, a_{t+1})\}_{i=1}^{N_p}$
and uses them to approximate the marginal differential entropy, assuming a marginal multivariate Gaussian distribution:
\begin{equation}
 H(\textbf{y}_{t+1} | a_{t+1}, \hat{\bar{\textbf{y}}}_{0:t}) \propto \frac{1}{2} \log( |\mathbf{\Sigma}_{\textbf{y}_{t+1}}(a_{t+1}, \hat{\bar{\textbf{y}}}_{0:t})|)~,
\end{equation}
where $\mathbf{\Sigma}_{\textbf{y}_{t+1}}(a_{t+1}, \hat{\bar{\textbf{y}}}_{0:t})$ is the covariance of the marginal  $p(\textbf{y}_{t+1} | a_{t+1}, \hat{\bar{\textbf{y}}}_{0:t})$, which we approximate from the samples $\{\textbf{y}_{t+1}^{i}|a_{t+1}\}_{i=1}^{N_p}$.

\subsection{Heart rate estimation}
At each iteration, the posterior mean is used as a point estimate of the angular position of the fetal heart: $x_t^{*}=\mathbb{E}_{q(x_{0:t}|\hat{\bar{\textbf{y}}}_{0:t})}(x_{t}) = \sum_{i=1}^{N_{p}} w_{t}^{i}x^{i}_{t}$. Its range $r_t^{*}$ is derived by searching for the maximum point in the power Doppler image along $x_t^{*}$. The one-dimensional Doppler signal at the estimated coordinates $\{x_t^{*}, r_t^{*}\}$ is processed by an (open-source) autocorrelation-based method \cite{valderrama2019open} to compute the HR.

\section{Experiments}
\label{sec:experiments}

\subsection{Validation Setup}

The performance of the proposed framework was assessed \textit{in-silico} and \textit{in-vitro}, comparing the performance of the closed-loop system with an actively steered focused beam with that of an open-loop system having a non-adaptive, fixed, ultrasound beam steered at 0~rad. 

For the \textit{in-silico} analysis, we used the Verasonics ultrasound simulator software \cite{cigier2022simus}. 
We model the beating heart through a set of point scatters distributed within a 2-cm-diameter disk at a depth of 13~cm. The scatter points move concentrically, following a sinusoidal motion having an amplitude comparable with fetal cardiac wall displacements and a frequency that yields a realistic HR. Fetal displacement relative to the ultrasound transducer is simulated by moving the scatter points laterally. We set a maximum relative displacement of 9~cm, which, given the target depth, corresponds to an angular displacement of 0.6~radians. We simulated static tissue scattering by adding a random distribution of point scatters of low reflectivity in the background. Imaging was simulated using an s5-1 (Philips) probe geometry and transducer parameters.

\begin{figure}[t!]
\centering
\includegraphics[width=0.45\textwidth]{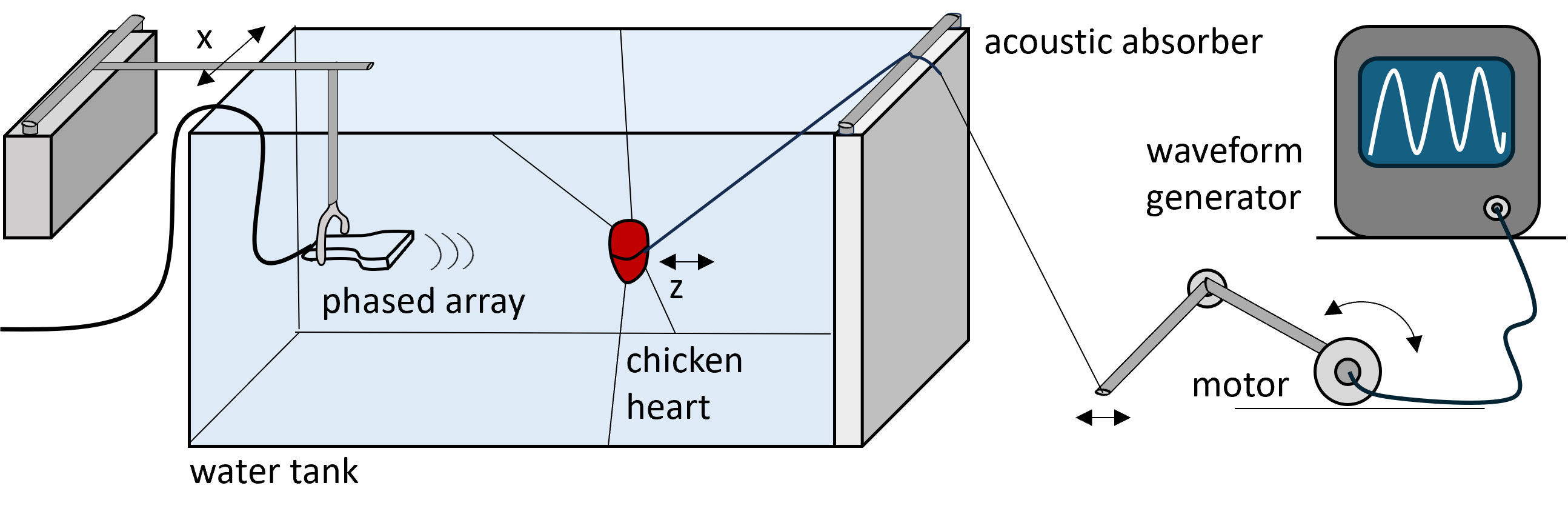}
\caption[Experimental setup.]{Experimental setup, including a submerged Philips s5-1 phased array probe manipulated by a translation stage and a suspended chicken heart attached to an axial motion generator.}
\label{fig:setup}
\end{figure}

For the \textit{in-vitro} analysis, we built an experimental setup based the one proposed by Hamelmann \textit{et al.} \cite{hamelmann2019fetal}. As displayed in Fig.~\ref{fig:setup}, a chicken heart moves rhythmically along the axial-direction through a chain of threads driven by a waveform generator, and it is imaged by a submerged s5-1 (Philips) phased array. To mimic the relative displacement between the fetal heart and the transducer, the phased array is mounted on a translation stage, which in our experiments was programmed to displace laterally according to a sinusoidal wave. The maximum displacement with respect to the ultrasound transducer was  8~cm, leading to an angular range of $\pm 0.5$~rad. The distance between the phased array and the chicken heart was around 13~cm. We added cornstarch (Maizena) to the water tank to introduce some background scattering.

For both \textit{in-silico} and \textit{in-vitro} experiments, the focused beam had a fixed focal depth of 13~cm and steering angle between $\pm0.7~$rad. The pulse repetition frequency was set to 1.5~kHz and the ensemble length to 1~s, which was needed to cover the typical fetal HR range of 110~bpm-160~bpm. The set of beamsteering actions $\boldsymbol{\Theta}^{tx}$ considered by the agent spans 21 angles between -0.7~rad and +0.7~rad.

\subsection{Results}
\begin{figure*}[t!]
    \centering
    \begin{subfigure}{.32\textwidth}
        \centering
        \includegraphics[width=\textwidth]{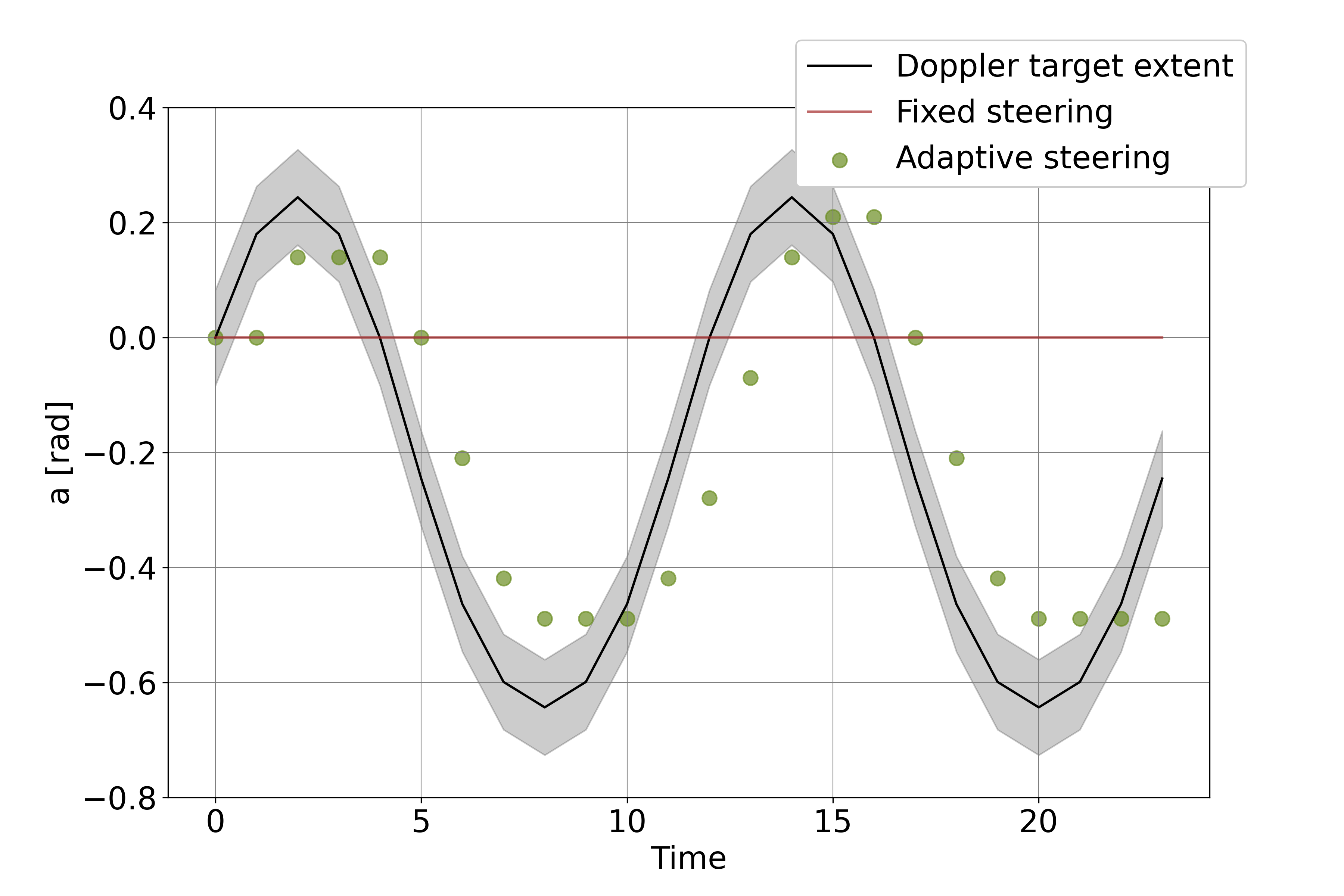}
        \caption{Steering Angle Selection}
    \end{subfigure}%
    ~ 
    \begin{subfigure}{.32\textwidth}
        \centering
        \includegraphics[width=\textwidth]{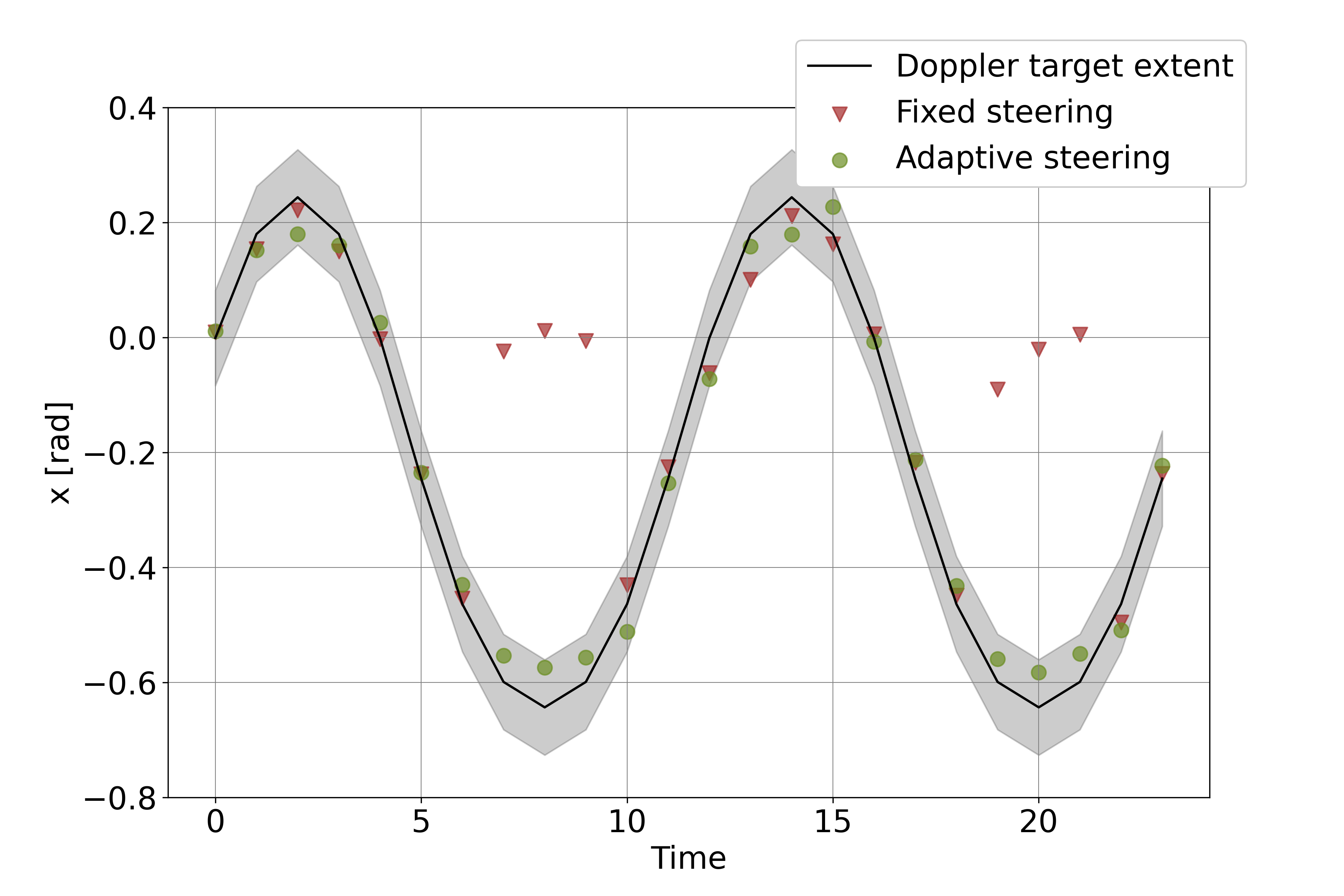}
        \caption{Angular Position Tracking}
    \end{subfigure}
    ~ 
    \begin{subfigure}{.32\textwidth}
        \centering
        \includegraphics[width=\textwidth]{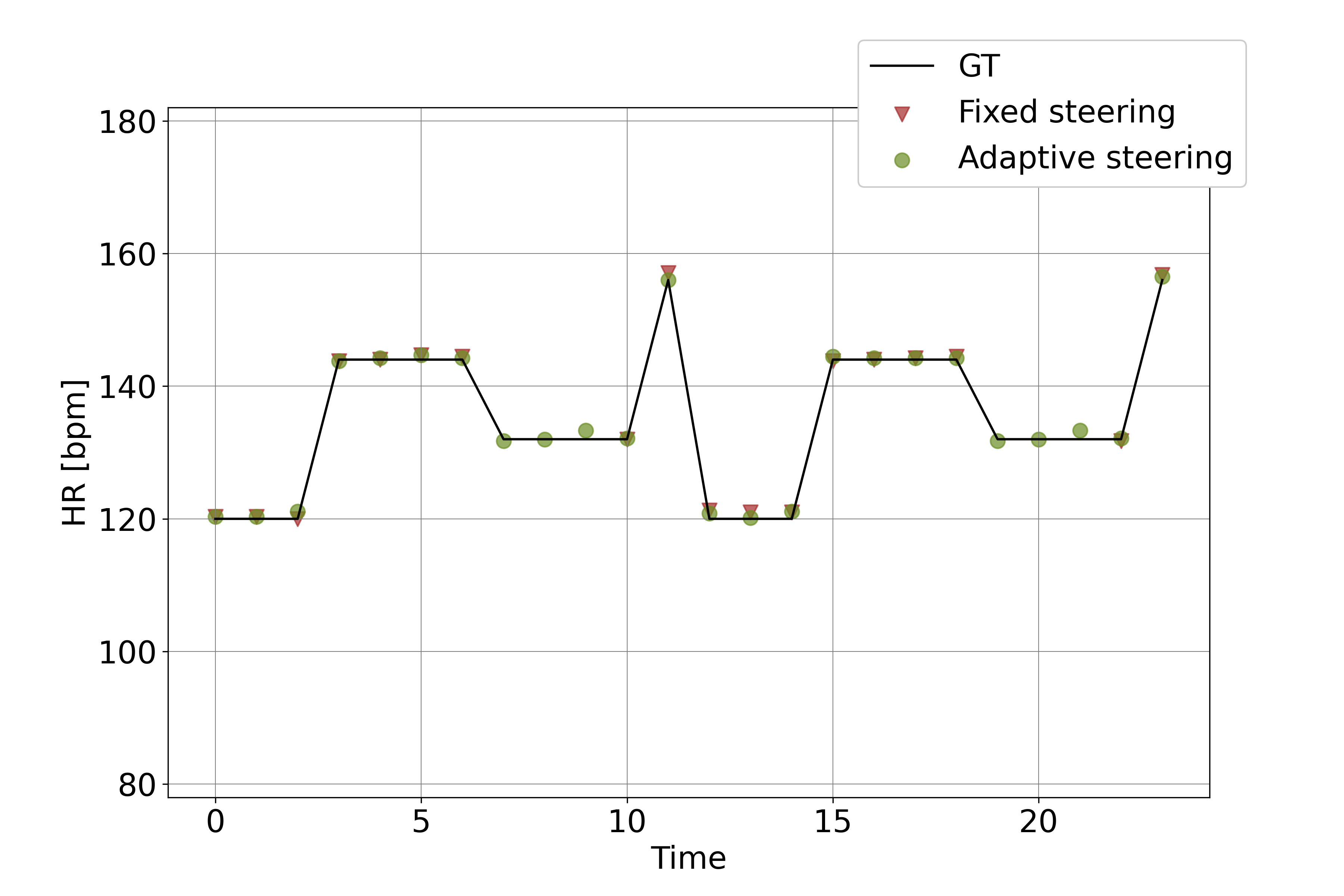}
        \caption{Heart Rate Estimate}
    \end{subfigure}
    \caption{\textit{In-silico} offline experiment comparing the performance of the proposed closed-loop system with adaptive focused beam and an open-system with a focused beam fixed at 0~rad at an SNR of 10~dB. (a) Selected steering angle compared to the Doppler target extent. (b) Angular position tracking over time. (c) Heart rate (HR) estimate over time compared to ground truth (GT). Note how the HR estimation fails when the Doppler target moves out of the fixed transmit beam.}
    \label{fig:1}
\end{figure*}

A representative example of the \textit{in-silico} offline experiments is shown in Fig.~\ref{fig:1}, where we compared the performance of the proposed adaptively-steered focused beam (green) with an ultrasound beam fixed at 0~rad (red), \textit{i.e. the best a-priori guess}. The adaptive strategy outperforms the fixed beam when the target displaces substantially from the transducer location. As visualized in Fig.~\ref{fig:1}a, the beam steering angle that maximizes information gain $ I(x_{t+1}, \textbf{y}_{t+1} | a_{t+1}, \hat{\bar{\textbf{y}}}_{0:t})$ effectively tracks the target angular position. By steering the beam close to the target, the target signal benefits from both transmit and receive beamforming gain, ensuring a strong signal and consistent detection. In contrast, with a fixed steering angle, the target eventually drifts outside the beam, causing the target Doppler signal to fall below the noise level, leading to the tracking errors visible in Fig.~\ref{fig:1}b and an inaccurate HR estimation displayed in Fig.~\ref{fig:1}c.

\begin{figure}[t!]
\centering
\includegraphics[width=0.48\textwidth]{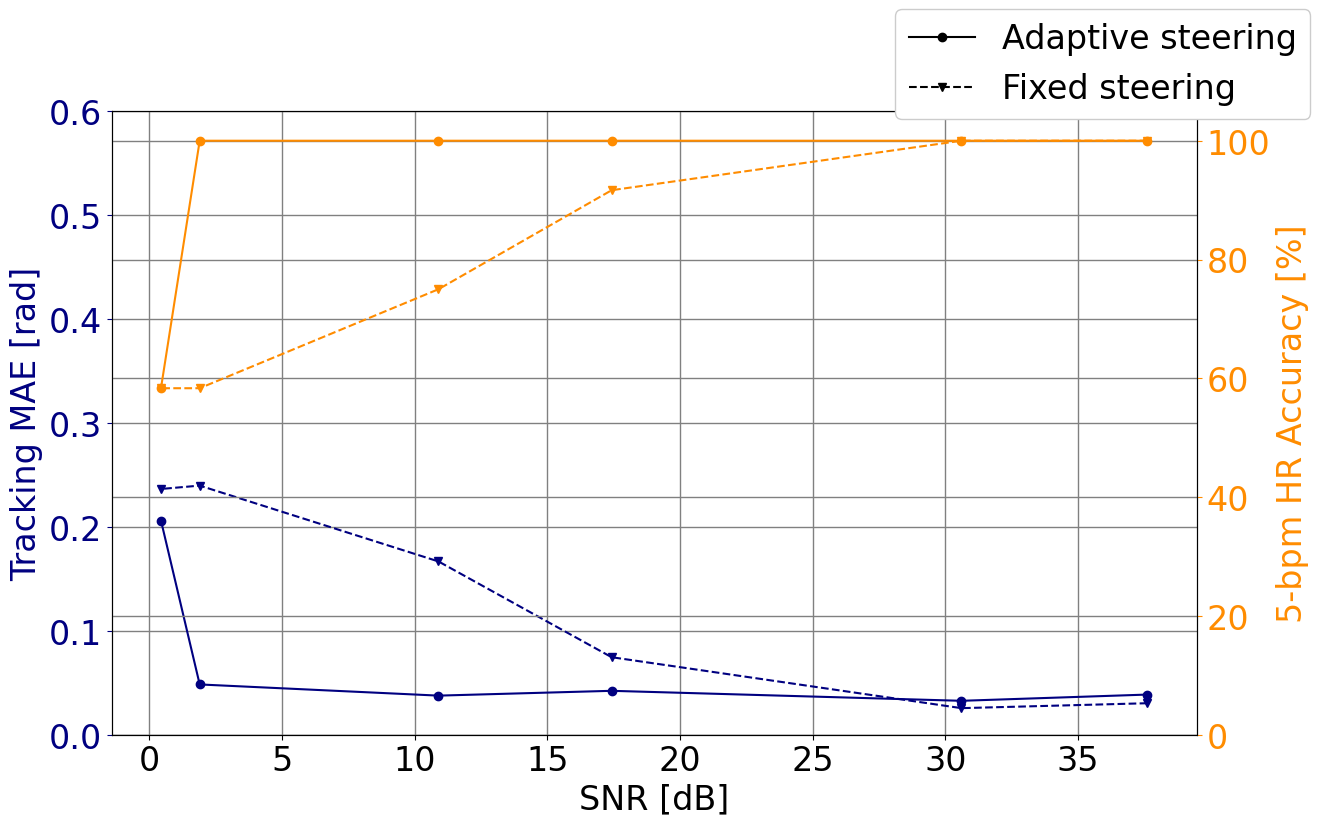}
\caption[]{Tracking mean absolute error (MAE) and heart rate (HR) accuracy ($\mathrm{|GT - HR|}$ $\leq$ 5~bpm) as a function of the power Doppler signal-to-noise ratio (SNR), for the fixed and adaptive beam steering policies. }
\label{fig:2}
\end{figure}

We then quantified the performance of the proposed approach for different SNR conditions. To that end, we added various levels of white Gaussian noise to the measured channel data. Fig.~\ref{fig:2} shows both the tracking MAE in radians and the accuracy of fetal HR estimation with a tolerance of 5~bpm, against the power Doppler SNR at the target depth without any transmit focusing gain. The latter is computed by transmitting with a single element and performing full coherent receive focusing, clutter filtering, and power Doppler processing.
The proposed approach with an adaptively-steered focused beam (continuous line) maintains excellent tracking performance and accurate HR estimation for a wider range of SNR levels, down to 0 dB. The fixed beam strategy (dashed line), instead, is unable to properly track the target, dropping its Doppler signal power below the noise level by not benefiting from an optimized transmit beamforming gain, in turn resulting in an inaccurate downstream HR estimate.

As for the \textit{in-vitro} online experiments, we found that the proposed closed-loop system can accurately track the position of the chicken heart even when the Doppler target is continuously moving, and the agent needs to infer its position and the optimal steering angle \textit{on-the-fly}. Over a target trajectory spanning three periods of the sinusoidal waveform controlling the probe's lateral motion, we measured a tracking mean absolute error of 0.03 rad and a 5-bpm HR accuracy ($\mathrm{|GT - HR|}$ $\leq$ 5 bpm) of 82\%. In terms of inference time, the posterior state estimate and action selection are completed in less than 1~s using a non-optimized implementation running on a CPU. 

\section{Discussion}
\label{sec:discussion}

\noindent This work presents a closed-loop cognitive ultrasound system designed to improve the performance of current fetal Doppler ultrasound systems. We propose to add, upstream the fetal HR estimation task, an agent that sequentially selects the optimal focused transmit beam to accurately track the fetal heart angular position.

Our findings indicate that by adjusting the steering angle of a focused beam to maximize expected information gain, the closed-loop system demonstrates robustness against both noise and displacement, effectively addressing the trade-offs faced by most conventional non-adaptive systems.

We also found that the positional information maximization objective aligns with the downstream fetal HR estimation objective. Although the agent's primary objective is to select the beam steering angle that improves tracking of the Doppler target, this adaptive beam steering also indirectly benefits the downstream HR estimation task.  In fact, to minimize uncertainty in the target's position, the agent tends to steer the ultrasound beam towards the target. By focusing the beam on the target, the Doppler signal benefits from both transmit and receive beamforming, leading to a higher Doppler SNR and a more accurate HR estimate.

Our generative model is relatively simple, and only loosely approximates the true physical data generating process. Yet, the model seems to be sufficiently accurate for facilitating adequate posterior inference and action selection in both \textit{in-silico} and \textit{in-vitro} experiments. For more complex environments and targets the modelling may need to be expanded, e.g. by learning parts of the model from data (i.e. deep generative modeling). Nevertheless, we here show that closed-loop systems equipped with simple models can already significantly enhance the overall system performance with favourable robustness.


\section{Conclusion}
\label{sec:conclusion}
In this work, we showcased the potential of closed-loop ultrasound in boosting the capabilities of traditional ultrasound imaging. We focused on improving
the robustness of conventional fetal HR monitoring
systems against fetal and maternal movements. To this end,
we augmented a fetal Doppler ultrasound system with a rational
agent that autonomously steers a focused transmit beam to
accurately track the fetal heart position. Our results show that
by dynamically adjusting the steering angle, this cognitive
system delivers accurate HR estimates in the presence of strong noise and substantial target movement. 


\bibliographystyle{ieeetr}
\bibliography{references}

\end{document}